\documentclass[prb,twocolumn,superscriptaddress,floatfix]{revtex4}
\usepackage{pictex}
\usepackage{psfig}
\usepackage{epsf}
\usepackage{graphicx}
\def\k{{\bf{k}}}
\def\grad{\nabla}

\begin{document}
\title{ARPES Spectra of the Hubbard model}
\author{Th.\ A.\ Maier}
\affiliation{Department of Physics,University of Cincinnati, Cincinnati 
OH 45221, USA}
\author{Th.\ Pruschke}
\affiliation{Center for electronic correlations and magnetism,
Theoretical Physics III, Institute for Physics, University of Augsburg, 
86135 Augsburg, Germany}
\author{M.\ Jarrell}
\affiliation{Department of Physics,University of Cincinnati, Cincinnati 
OH 45221, USA}

\begin{abstract}
We discuss spectra calculated for the 2D Hubbard model in the intermediate
coupling regime with the dynamical cluster approximation, which is a
non-perturbative approach. We find a crossover
from a normal Fermi liquid with a Fermi surface closed around the Brillouin
zone center at large doping to a non-Fermi liquid for small doping. The 
crossover is signalled by a splitting of the Fermi surface around the $X$
point of the $2D$ Brillouin zone, which eventually leads to a hole-like Fermi
surface closed around the point $M=(\pi,\pi)$. The topology of the Fermi
surface at low doping indicates a violation of Luttinger's theorem. We discuss
different ways of presenting the spectral data to extract information about
the Fermi surface.  A comparison to recent experiments will be presented.
\end{abstract}
\pacs{}
\maketitle              

\section*{Introduction}

The rich phenomenology of high-$T_c$ superconductors \cite{htc_review} has 
stimulated strong experimental and theoretical interest in the field of 
strongly correlated electron systems.  Apart from the anomalously high
transition temperatures, these compounds are also of interest due to their 
unusual normal state properties.
Most of these anomalous properties are found in spectra 
and transport quantities, i.e.\ are intimately linked to the dynamics of 
the electronic degrees of freedom. Thus, much of the experimental and 
theoretical effort has concentrated on the development of an understanding 
of the single-particle dynamics.  Among the fundamental and controversial
questions are whether the cuprates can be described as a Fermi liquid or 
not and what shape and volume a possible Fermi surface will have.

In this connection, one of the most informative experimental probes of the 
cuprates has become the Angle-Resolved Photoemission Spectroscopy(ARPES). 
The development in this field, again largely stimulated by the interest in 
the physics of high-$T_c$, has led to a tremendous increase of the angular 
and energy resolution \cite{Borisenko,Damascelli}.  ARPES is now able to 
access the low-energy behavior of single-particle spectra and especially 
the shape and topology of the cuprate Fermi surface.  This has lead to the
discovery of the shadow bands\cite{kampf,htc_review} and the pseudo-gap 
formation in the underdoped cuprates\cite{htc_review}.  Recently, the 
single-particle self energy was extracted from the ARPES data, showing 
extremely interesting behavior especially close to $(\pi/2,\pi/2)$ on 
the Fermi surface\cite{ARPES_SIGMA}.

A large number of the more recent ARPES experiments have concentrated on 
Fermi surface mapping. Some of these experiments seem to indicate that, at
least for LSCO, the Fermi surface switches from being hole-like, centered 
at $M=(\pi,\pi)$, for low doping to being electron-like, centered at the 
zone center $\Gamma=(0,0)$ for high doping with a volume consistent with 
the doping level\cite{Damascelli,Fujimori,Yoshida}.  In other results, 
particularly for bilayer compounds like YBCO or Bi2212\cite{Kordyuk,Damascelli},
the Fermi surface seems to remain hole-like independent of the doping 
level. Especially for those latter compounds, further complications 
in the experiments arise from superstructures due to umklapp scattering 
and the possibility of bilayer splittings in YBCO and Bi2212\cite{Bogdanov,
Damascelli}. Furthermore, in the underdoped regime one can observe 
additional changes in the Fermi surface topology that are frequently
brought in connection with the possibility of stripe formation in 
LSCO\cite{Damascelli}.

The presence of a large Fermi surface has been taken as a validation of 
Luttinger's theorem; however, other results find that the Fermi surface 
volume, at least in the underdoped regime, is too small \cite{Kordyuk}, 
which would point towards a violation of Luttinger's theorem. Furthermore, 
recent experiments seem to indicate that the low doping Fermi surface near 
$X=(\pi,0)$ actually bifurcates into two parts, one electron-like and one 
hole-like\cite{Bifurcation}.  This splitting has been interpreted in terms 
of strong interlayer coupling\cite{Bogdanov}, but could equally well result 
from shadow Fermi surface formation due to coupling of the electrons to 
strong antiferromagnetic fluctuations around the $X$ points\cite{Kordyuk}.

Thus, a consistent experimental picture concerning both shape and volume 
of the Fermi surface is at present not available and a theoretical 
investigation of the generic features to be expected based on a model 
calculation is necessary.

Early in the theoretical investigation of the high-$T_c$ cuprates it was 
realized that the 2D Hubbard model
\begin{equation}\label{equ:1}
H=\sum\limits_{i,j,\sigma}
t_{ij}c^\dagger_{i\sigma}c^{\phantom{\dagger}}_{j\sigma}+\frac{U}{2}
\sum\limits_{i\sigma}c^{\dagger}_{i\sigma}c^{\phantom{\dagger}}_{i\sigma}
c^{\dagger}_{i\bar{\sigma}}c^{\phantom{\dagger}}_{i\bar{\sigma}}
\end{equation}
in the intermediate coupling regime, or closely related models like the 
$t$-$J$ model probably capture the essential physics \cite{Anderson}.  
In the wake of this conjecture, a huge effort has been directed to the
study of these models\cite{dagotto}. 
There is now a general consensus that the appropriate parameter regime
for the cuprates is the intermediate coupling regime where the Coulomb 
parameter $U$ is roughly equal to the bandwidth.  However, this is the
most complicated regime of the model since both weak and strong coupling
perturbative approaches fail.  Exact diagonalization of small 
clusters \cite{dagotto} suffers from strong finite-size effects, often 
ruling out the reliable extraction of low-energy dynamics.
Conventional Quantum Monte Carlo for finite sized systems suffers from a 
severe minus sign problem in this parameter regime.  The resulting data
is of insufficient quality to allow for reliable calculations of dynamic 
quantities at low enough temperatures.  High-temperature series has
provided some of the most informative results for the Fermi surface 
topology, but it does not yield spectra, and, so far, only results for 
the t-J model are available\cite{puttika}.

Thus, also from a theoretical point of view, the question of whether a 
Fermi surface does actually exist and what its topology is still is a matter 
of debate. In that connection it is of special interest that some 
experiments  indicate a violation of Luttinger's theorem; if true, any
theory, such as FLEX\cite{flex}, based on a weak-coupling expansion 
around the non-interacting limit would be inadequate.

Thus a treatment within a non-perturbative scheme
clearly is desirable.
In this paper we therefore use the recently developed dynamical cluster approximation 
(DCA) \cite{DCA_hettler,DCA_maier,DCA_huscroft, DCA_moukouri0,DCA_QMC_Jarrell} 
to study the low-energy behavior of the $2D$ Hubbard model in the 
intermediate coupling regime with nearest-neighbor hopping $t$ and on-site 
correlation $U$ equal to the band width $W$.  The DCA systematically
incorporates non-local corrections to local approximations like the 
dynamical mean field, by mapping the lattice onto a self-consistently
embedded cluster.  We solve the cluster problem using a combination of
quantum Monte Carlo (QMC) and the maximum entropy method to obtain dynamics.
This technique produces results in the thermodynamic limit and has 
a mild minus-sign problem\cite{DCA_QMC_Jarrell}.

The paper is organized as follows. The next section contains a brief
introduction to the DCA. The numerical results will be
presented in the third section followed by a discussion and summary.

\section*{Formalism}
A detailed discussion of the DCA formalism was already given in previous 
publications \cite{DCA_hettler,DCA_maier,DCA_huscroft,DCA_moukouri0,
DCA_QMC_Jarrell}.  The main assumption underlying the DCA is that the 
single-particle self-energy $\Sigma(\vec{k},z)$ is a slowly varying 
function of the momentum $\vec{k}$ and can be approximated by a constant 
within each of a set of cells centered at a corresponding set of
momenta $\vec{K}$ in the first Brillouin zone\cite{DCA_hettler}. 
The single-particle lattice Green functions are then coarse-grained
or averaged within these cells, and used to calculate the lattice 
self energy and other irreducible quantities. Within this approximation, 
one can set up a self-consistency cycle similar to the one in the 
dynamical mean-field theory (DMFT)\cite{dmft}. However, in contrast 
to the DMFT, where only local correlations are taken into account, 
the DCA includes non-local dynamical correlations. The length scales 
of these non-local correlations can be varied systematically from short
ranged to long ranged by increasing the number of coarse-graining cells.
The DCA collapses to the DMFT if one represents the Brillouin zone
by one cell only, thus setting the characteristic length scale to zero.

By construction, the DCA preserves the translational and point group 
symmetry of the lattice. Comparisons to other extensions of the DMFT 
developed during the past years\cite{lichtenstein,kotliar} either show 
that these are identical to the DCA or converge more slowly as function 
of cluster size \cite{DCA_MCPA_COMP}.

For the impurity problem of the DMFT a large set of reliable numerical techniques
has been developed over the past ten years\cite{DMFA_Jarrell1,dmft,nrg}. We 
have employed Quantum Monte Carlo (QMC) and the Non-Crossing Approximation 
(NCA) \cite{DCA_maier} as non-perturbative (in $U$) methods to solve the 
cluster problem of the DCA.  For cluster sizes larger than $N_c=4$ however, only the 
QMC technique is presently available for the DCA.  The cluster problem is 
solved using the Hirsch-Fye impurity algorithm \cite{hirsch_fye} modified 
to simulate an embedded cluster\cite{DCA_QMC_Jarrell}. Note that this problem 
is computationally much more difficult than that encountered in either
the DMFA or in finite-sized simulations, since the block diagonal 
structure in space-time occurring in conventional finite-system simulations 
is not present here. This increase in computation time is, however, partially 
compensated by a rather mild minus-sign problem, even for comparatively 
large values of $U$ and small temperatures\cite{DCA_QMC_Jarrell}. From the 
QMC data, the spectra are obtained by analytic continuation with the maximum 
entropy method\cite{JARRELLandGUB}.
Finally, the self energy is interpolated on to the full Brillouin zone using
Akima splines, which is a sensible step as long as the assumption of a slow
variation in $\k$-space is valid. Note that it is very important to
interpolate irreducible quantities like the self energy and {\em not}\/ for
example the cluster Green function itself.

\section*{Results}
\begin{figure}[ht]
\includegraphics[width=3.0in]{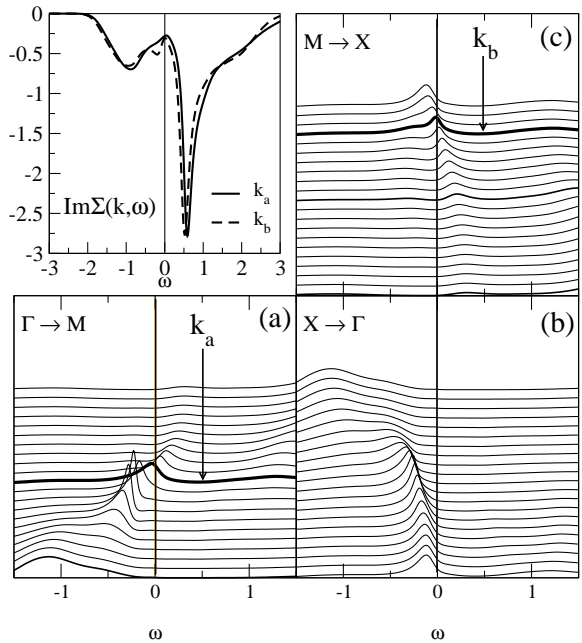}
\caption{ (a)--(c) The single-particle spectrum $A(\k,\omega)$ for $U=2$eV, 
$T=\frac{1}{30}$eV, $\delta=0.05$, $t_\perp=0$ and $N_c=16$ along certain high 
symmetry directions (spectra for different $\bf k$ are shifted along the 
y-axis). The colored lines in Figs.~(a) and (c) indicate the spectra which 
cross the Fermi energy with a peak closest to $\omega=0$.  In (b), no 
such peak is found which crosses the Fermi energy.  (d) the imaginary 
part of the self energy versus frequency at the Fermi surface 
crossing found in (a) and (c).
}
\label{Akw_N0.95}
\end{figure}
For a proper description of the CuO$_2$ planes of the high-$T_c$ cuprates
within the Hubbard model (\ref{equ:1}) it is generally accepted that the
tight-binding dispersion has the form
\begin{equation}\label{equ:2}
t_{\vec k}=
\begin{array}[t]{l}\displaystyle
-2t\left(\cos(k_x)+\cos(k_y)\right)\\[5mm]
\displaystyle
-4t'\cos(k_x)\cos(k_y)
\end{array}
\end{equation}
with a nearest neighbor hopping amplitude $t>0$ and a next-nearest neighbor
hopping amplitude $t'$, which in principle can have any sign. From 
bandstructure calculations and the general form of the measured Fermi 
surface, especially in the overdoped regime, conventionally a negative 
$t'$ is inferred\cite{bandstructure}. Such a negative $t'$ would naturally
lead to a Fermi surface closed around the Brillouin zone corner $M$.  The
interesting  question, however, is whether the Fermi surface closed around
the $M$ point observed experimentally in the low-doping region is a mere
bandstructure effect or induced by correlations. Weak coupling treatments
of the $2D$ Hubbard model indicate that such a change of the shape,
but not the volume, of the
Fermi surface due to the interactions is indeed possible\cite{zlatic,flex}.
Thus, to obtain insight into the effects of correlations on the structure
of the Fermi surface and distinguish them from pure bandstructure effects,
we concentrate on the case $t'=0$ in this paper.

In the following we set $t=1/4$eV in accordance with typical values extracted
from the experiments and bandstructure and choose $U=W=2$eV. This value of
$U$ is sufficiently large that for $N_c\geq 4$ a Mott gap is present in the
half-filled  model \cite{DCA_moukouri1}. We performed our simulations at a 
range of temperatures, but will present results for $T=0.033$eV only, which 
is roughly room temperature.  A pseudogap due to short-ranged spin 
correlations is also present in the weakly doped model for slightly lower 
temperatures \cite{DCA_QMC_Jarrell}.
                    
\begin{figure}[ht]
\includegraphics[width=3.0in]{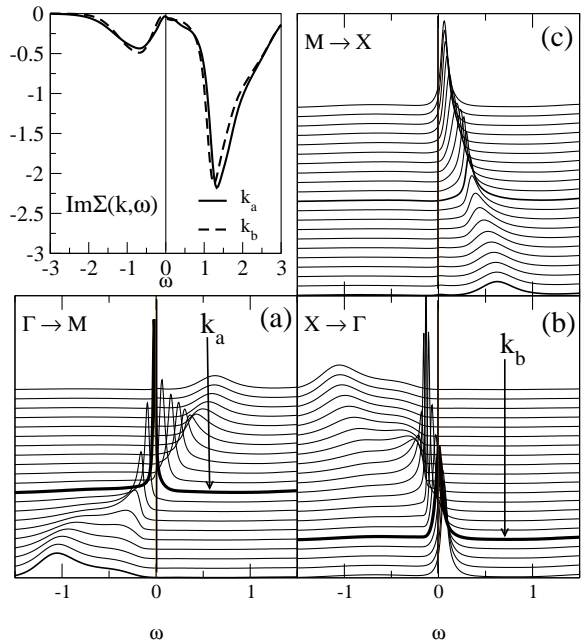}
\caption{ (a)--(c) The single-particle spectrum $A(\k,\omega)$ $\delta=0.20$
along certain high symmetry directions. Other parameters as in 
Fig.~\ref{Akw_N0.95}. The colored lines in Figs.~(a) and (b) indicate the 
spectra which cross the Fermi energy with a peak closest to $\omega=0$.  
In (c), no such peak is found which crosses the Fermi energy.  (d) the 
imaginary part of the self energy versus frequency at the Fermi 
surface crossing found in (a) and (b).
}
\label{Akw_N0.80}
\end{figure}

The single-particle spectra for certain high symmetry 
directions are plotted in Figs.~\ref{Akw_N0.95} and \ref{Akw_N0.80} for
$n=0.95$  and $n=0.80$, respectively.  We use the standard convention
to identify the high symmetry points in the zone, $\Gamma=(0,0)$,
$M=(\pi,\pi)$ and $X=(\pi,0)$.  For $n=0.95$ the peak in the spectrum
crosses  the Fermi energy along the $\Gamma\to M$ and $M\to X$
directions, while for $n=0.80$ the second crossing appears along $X\to\Gamma$.
The imaginary part of the self energy at these crossing points is plotted
versus frequency in panel (d) of Figs.~\ref{Akw_N0.95} and \ref{Akw_N0.80}.  

One very interesting feature of the spectra at low doping, 
Fig.~\ref{Akw_N0.95}, is that the peak near ($\pi/2,\pi/2$) 
broadens dramatically before crossing the Fermi energy.
Near $X$, on the other hand, one does not observe any dramatic change
in the spectrum when crossing the Fermi energy.
This indicates that near ($\pi/2,\pi/2)$ hole-like 
quasi-particle (QP) excitations with $k < k_F$ appear to have
longer  lifetimes than electronic excitations with $k > k_F$. This asymmetry
between particles and holes near the Fermi surface is a strong indication of
non-Fermi liquid (NFL) behavior, at least along the $\Gamma$-$M$ direction.
Note that this is at least qualitatively in agreement with experimental
ARPES spectra\cite{Damascelli,Ino}. Although these experiments on LSCO had to
be performed below $T_c$ in the underdoped sample\cite{Ino}, the authors
infer from the general behavior of the spectra, that a peak crossing the
Fermi energy can be found along $M\to X$, while a rather broad structure
is seen along $\Gamma\to M$ around $(\pi/2,\pi/2)$. It would clearly
be interesting to have experimental ARPES data above $T_c$ available for a
more founded comparison.

It is also quite instructive to look at the imaginary part of the self-energy
at the $\bf k$-points where the peak in the spectrum crosses the  Fermi energy
(Fig.~\ref{Akw_N0.95}(d)). In particular at the crossing point $\k_b$ close to
$(\pi,0)$, $\Im m \Sigma(\k_b,\omega)$ shows a striking asymmetry in the
low-frequency regime as compared to $\Im m \Sigma(\k_a,\omega)$, where $\k_a$
is the crossing point close to $(\pi,0)$. In fact, $\Im m \Sigma(\k_b,\omega)$
starts to develop an additional feature which eventually leads to the
formation of a pseudo gap in the spectra along the $M\to X$
direction\cite{DCA_QMC_Jarrell}. In addition, one observes a rather large
residual scattering rate for both momenta $\vec k_a$ and $\vec k_b$.
This can either be taken as further evidence for NFL behavior or as signal
for the occurrence of a new very small low-energy scale\cite{Altmann}. 
Obviously, we cannot decide this question on the basis of the present data,
but would have to look at much lower temperatures. Unfortunately, this is not
possible at present.  Note that in the low-energy regime the self energy 
displays significant $k$-dependence. This clearly renders theories based 
on a local approximation like the DMFA inadequate at least for small doping.

At higher doping, Fig.~\ref{Akw_N0.80}, the peaks in the spectrum close
to the Fermi energy are far sharper. Here it makes sense to speak of
a conventional Fermi liquid and quasi particles again. As already mentioned, 
the Fermi energy crossings can be found along $\Gamma\to M$ and $X\to\Gamma$.
Again, this is in qualitative accordance with ARPES experiments for strongly
overdoped LSCO\cite{Damascelli,Ino}, although these experiments still find
rather broad structures even in heavily overdoped samples.
In addition, there is no evidence for particle-hole asymmetry in our data.
Especially at $(\pi/2,\pi/2)$ the structure crossing the Fermi level 
appears to be rather symmetric with  respect to the crossing point.

\begin{figure}[ht]
\begin{center}
\beginpicture
\setcoordinatesystem units <3in,3in>
\setplotarea x from 0 to 1, y from 0 to 1
\axis bottom invisible label {$k_x$} ticks withvalues $0$ $\pi$ / at 0 1 / /
\axis left invisible label {$k_y$} ticks withvalues {} $\pi$ / at 0 1 / /
\put {\includegraphics[width=2.95in]{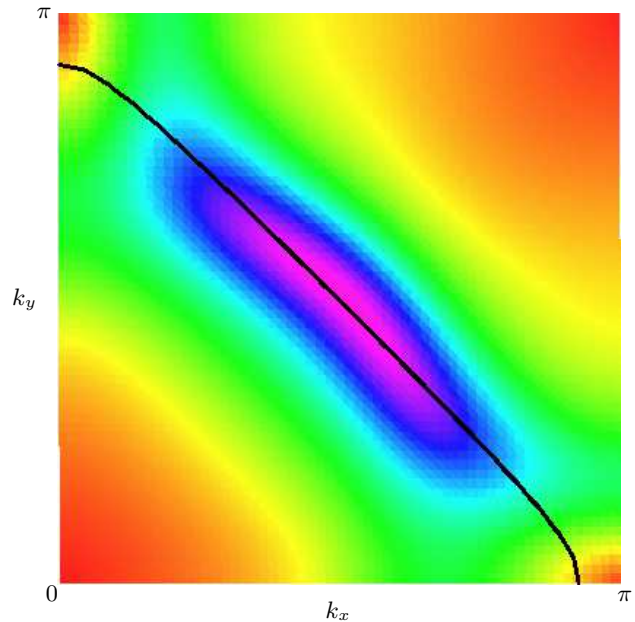}} at 0.501 0.501
\endpicture
\end{center}
\caption{Gradient of the distribution function $|\grad n(\k)|$ for $U=2$eV, 
$T=\frac{1}{30}$eV, $\delta=0.05$, $t_\perp=0$ and $N_c=16$. Shown is the upper
right quadrant of the first BZ.
The violet (red),
represents regions of high (low) values.  The blue and violet
regions map out the Fermi surface. The non-interacting Fermi surface
is shown by the black line. The interacting Fermi surface is rather hard to
define. Note the strong bifurcation around the $X$ points.
}
\label{095_lowT_gradnk}
\end{figure}
\begin{figure}[ht]
\begin{center}
\beginpicture
\setcoordinatesystem units <3in,3in>
\setplotarea x from 0 to 1, y from 0 to 1
\axis bottom invisible label {$k_x$} ticks withvalues $0$ $\pi$ / at 0 1 / /
\axis left invisible label {$k_y$} ticks withvalues {} $\pi$ / at 0 1 / /
\put {\includegraphics[width=2.95in]{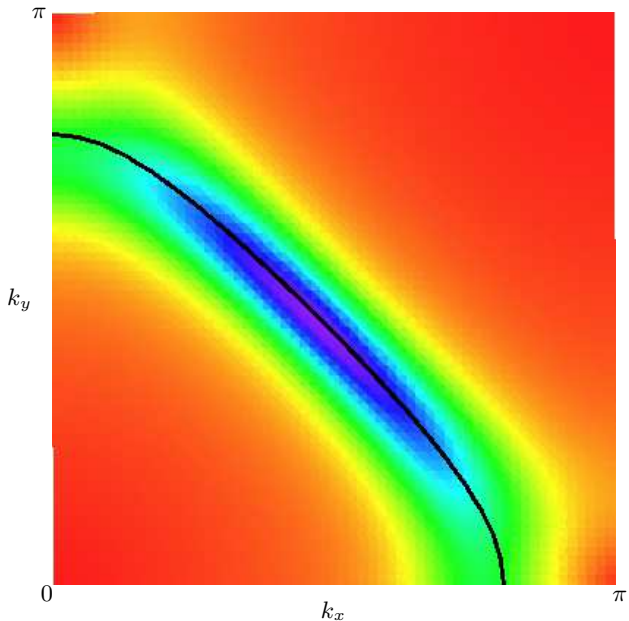}} at 0.501 0.501
\endpicture
\end{center}
\caption{$|\grad n(\k)|$ for $U=2$eV, $T=\frac{1}{30}$eV, 
$\delta=0.20$, $t_\perp=0$ and $N_c=16$. The color scheme is the same 
as in Fig.~\ref{095_lowT_gradnk}.  The non-interacting and interacting Fermi 
surfaces coincide, in accord with Luttinger's theorem.
}
\label{080_lowT_gradnk}
\end{figure}
The imaginary part of the self energy, shown in
Fig.~\ref{Akw_N0.80}(d), has a broad maximum at $\omega=0$ with a very
small residual scattering rate and changes little as $\k$ moves along the
Fermi surface. This weak dependence on $\k$ is an indication that approximations
like the DMFA should be accurate here, i.e.\ that there is little effect of
non-local correlations. All indications are that for this doping regime
standard Fermi-liquid behavior has returned.

More evidence for NFL behavior can be seen in the shape of the Fermi surface. 
Theoretically, the Fermi surface can be defined in two different ways. First, 
the gradient of the electronic distribution function, $|\grad n(\k)|$, has a 
maximum at the Fermi surface. From a computational point of view this quantity 
is very convenient, because it does not require the calculation of dynamical 
properties. However, especially for a comparison with experimental Fermi 
surface mappings based on ARPES experiments, the approach via $|\grad n(\k)|$ 
probably is not adequate, mainly due to the unknown influence of matrix 
elements in the experimental spectra\cite{Borisenko}.

An analysis of the Fermi surface for the $t$-$J$
model based on a study of $|\grad n(\k)|$ has been performed recently within
a high-temperature expansion\cite{puttika} and been considered as clear
evidence for a violation of Luttinger's theorem and the possible formation
of a non-Fermi liquid at small doping.
Our results for that quantity are collected in Figs.~\ref{095_lowT_gradnk}
and \ref{080_lowT_gradnk}, where $|\grad n(\k)|$  for the upper right quadrant
of the first Brillouin zone is shown in a density plot for $n=0.95$ and
$n=0.8$, respectively. Regions of large values of $|\grad n(\k)|$ are colored
in blue and violet, regions of small values in red. For comparison the
Fermi surface for the non-interacting system is included (black line).
For small doping, Fig.~\ref{095_lowT_gradnk}, $|\grad n(\k)|$ gives a
rather broad structure around $(\pi/2,\pi/2)$ rather following the
noninteracting Fermi surface. It is especially hard to define a Fermi surface
at all or extract a reliable estimate of the Fermi surface volume from these
results. Note also the strong bifurcation around the $X$ points.
We find that especially for small doping $|\grad n(\k)|$ generally
predicts that the Fermi surface bifurcates near $X$, with parts of the Fermi
surface along both $M\to X$ and $X\to \Gamma$. This is in striking contrast 
to the fact that the spectra in Fig.~\ref{Akw_N0.95} do not show a 
crossing along the direction $X \to \Gamma$, only a peak coming close to the
Fermi energy. Apparently, $|\grad n(\k)|$ is unable (at least at these
temperatures) to distinguish between a peak in the spectrum crossing or just
simply approaching the Fermi energy.

In accordance with the spectra in Fig.~\ref{Akw_N0.80}, the plot of $|\grad n(\vec k)|$
for large doping in Fig.~\ref{080_lowT_gradnk} shows a fairly well defined
Fermi surface that coincides with the Fermi surface of the non-interacting
system. Again, in contrast to the sharp peaks found in the spectra, 
$|\grad n(\vec k)|$ shows a substantial broadening, which in this case can
however be explained by the conventional temperature broadening of Fermi's
function.

An alternative way of mapping the Fermi surface, which also allows more direct
contact with ARPES experiments, is to make constant energy plots of the
single-particle spectra at the Fermi energy $A(\k,\omega=0)$.
\begin{figure}[ht]
\begin{center}
\beginpicture
\setcoordinatesystem units <3in,3in>
\setplotarea x from 0 to 1, y from 0 to 1
\axis bottom invisible label {$k_x$} ticks withvalues $0$ $\pi$ / at 0 1 / /
\axis left invisible label {$k_y$} ticks withvalues {} $\pi$ / at 0 1 / /
\put {\includegraphics[width=2.95in]{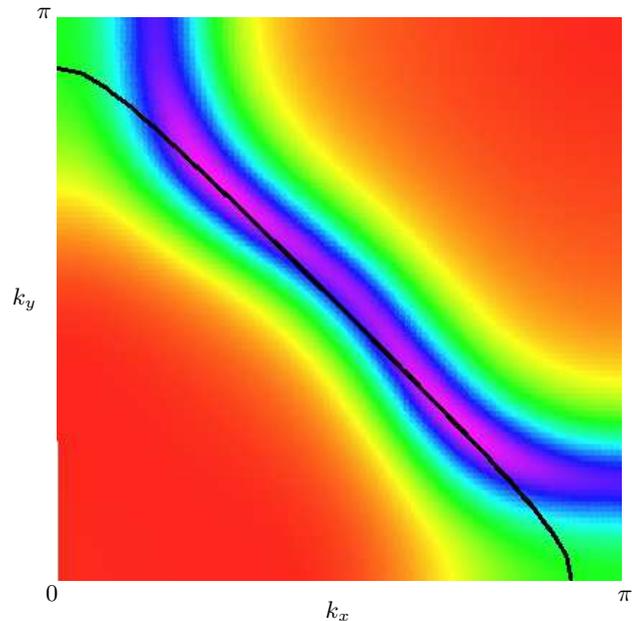}} at 0.501 0.501
\endpicture
\end{center}
\caption{Constant energy scans of $A(\k,\omega=0)$ for $U=2$eV, 
$T=\frac{1}{30}$eV, $\delta=0.05$, $t_\perp=0$ and $N_c=16$ in
the right upper quadrant of the first BZ. The violet (red),
represents regions of high (low) electronic density.  The blue and violet
regions map out the Fermi surface. The non-interacting Fermi surface
is represented by a black line. The interacting Fermi surface is hole-like,
centered at $(\pi,\pi)$, and encloses significantly more volume than
the non-interacting Fermi surface.  This indicates a violation of 
Luttinger's theorem.
}
\label{095_lowT_Nc}
\end{figure}
\begin{figure}[ht]
\begin{center}
\beginpicture
\setcoordinatesystem units <3in,3in>
\setplotarea x from 0 to 1, y from 0 to 1
\axis bottom invisible label {$k_x$} ticks withvalues $0$ $\pi$ / at 0 1 / /
\axis left invisible label {$k_y$} ticks withvalues {} $\pi$ / at 0 1 / /
\put {\includegraphics[width=2.95in]{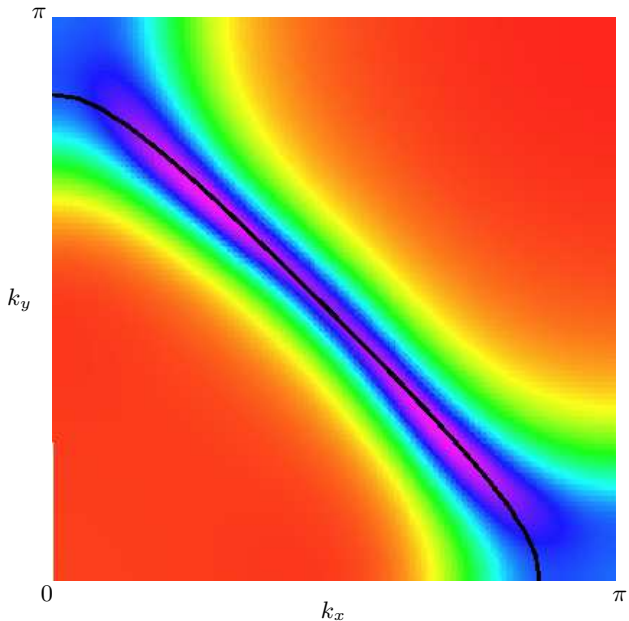}} at 0.501 0.501
\endpicture
\end{center}
\caption{Constant energy scans of $A(\k,\omega=0)$ for $\delta=0.1$, 
The other parameters and the color scheme are the same 
as in Fig.~\ref{095_lowT_Nc}.  The non-interacting and interacting Fermi 
surfaces start to coincide again, which can be interpreted as restoration
of Luttinger's theorem. Note however the strong bifurcation around the $X$
point.}
\label{090_lowT_Nc}
\end{figure}
\begin{figure}[ht]
\begin{center}
\beginpicture
\setcoordinatesystem units <3in,3in>
\setplotarea x from 0 to 1, y from 0 to 1
\axis bottom invisible label {$k_x$} ticks withvalues $0$ $\pi$ / at 0 1 / /
\axis left invisible label {$k_y$} ticks withvalues {} $\pi$ / at 0 1 / /
\put {\includegraphics[width=2.95in]{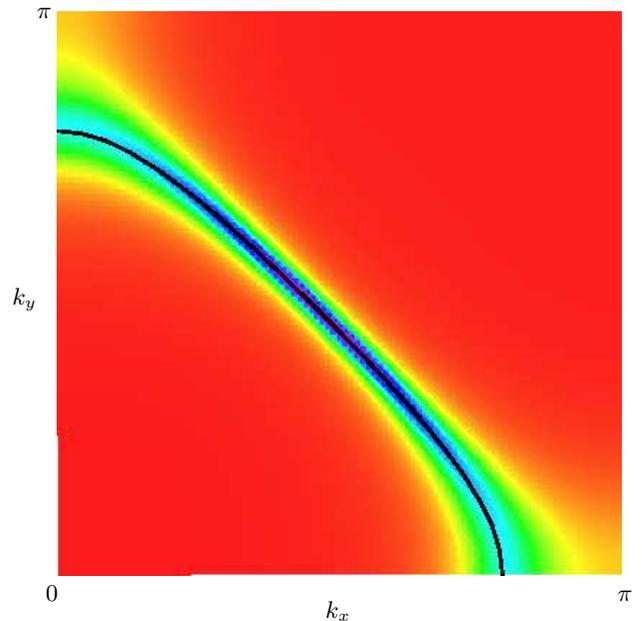}} at 0.501 0.510
\endpicture
\end{center}
\caption{Constant energy scans of $A(\k,\omega=0)$ for $\delta=0.2$.
The other parameters and the color scheme are the same 
as in Fig.~\ref{095_lowT_Nc}.  The non-interacting and interacting Fermi 
surfaces coincide, in accord with Luttinger's theorem.}
\label{080_lowT_Nc}
\end{figure}
The constant energy plots $A(\k,\omega=0)$ are shown in Figs.~\ref{095_lowT_Nc},
\ref{090_lowT_Nc} and \ref{080_lowT_Nc} for $n=0.95$, $n=0.90$  and $n=0.80$, respectively, for the
upper right quadrant of the first Brillouin zone.  The
regions of high density are colored in violet and low density
in red. The solid black lines as before represent the non-interacting Fermi
surface.
For $n=0.95$, the Fermi surface resulting from our calculations is hole-like,
centered around $(\pi,\pi)$, and encloses a volume larger than the non-interacting
Fermi surface, indicating a violation of Luttinger's theorem. In addition,
the shape of the Fermi surface close to $(\pi/2,\pi/2)$, especially the clear
shift above $(\pi/2,\pi/2)$ cannot be interpreted neither in terms of a simple
tight-binding band structure nor a weak-coupling theory\cite{zlatic}. 
Note also that in comparison to the $|\grad n(\k)|$ result, no apparent
electron like shadow band at $(\pi,0)$ can be seen. 

With increasing doping a bifurcation of the Fermi surface around $(\pi,0)$ 
starts to develop (see Fig.~\ref{090_lowT_Nc}), signalling the incipient 
crossover from a hole like Fermi surface to the expected electron
like at large doping. In addition, around $(\pi/2,\pi/2)$ the peak in the
spectrum follows more or less the noninteracting Fermi surface again. This
points towards a restoration of Luttinger's theorem. For $n=0.80$, the Fermi
surface is definitely electron like, centered at $(0,0)$, and has essentially
the same volume and shape as the non-interacting surface, indicating a return
to Fermi-liquid like behavior.
No apparent remnants of the hole-like Fermi surface at lower doping are left.
Interpreting these bifurcations seen most strongly around $n=0.90$ 
as shadow Fermi surfaces, we find a strong reduction of the weight in these
shadow features both for dopings less and larger than this ``optimal'' doping.
Interestingly, a similar behavior was recently observed in the analysis of
ARPES data for Bi2212\cite{Kordyuk}. However, in these experiments the Fermi
surface remained hole-like throughout the whole doping regime studied. Whether
this might be related to a finite $t'$ neglected in our present calculations
will be discussed elsewhere.

\section*{Summary and conclusions}
The increasing precision and quality of experimental ARPES spectra in recent 
years has led to a number of new results on the single-particle dynamics
of the high-$T_c$ cuprates, both partially resolving long-standing issues
and posing new questions and problems. Motivated by especially the interesting
observations concerning the changes of Fermi surface topology with doping,
we have studied the two-dimensional (2D) Hubbard model with nearest neighbor
hopping $t$ in the intermediate coupling regime (on-site correlation $U$ equal
to the bandwidth) at the temperature $T=0.033$eV. To study the
model in this most problematic parameter regime we used quantum Monte Carlo
within the  dynamical cluster approximation for the cluster size $N_c=16$.
Since this method allows for controlled and reliable calculations of 
low-energy features within a non-perturbative scheme and in the thermodynamic 
limit, fundamental problems in this field can be addressed.  These include
the single-particle spectral properties at low energies, especially possible 
deviations from Luttinger's theorem or the formation of non Fermi liquid 
states, and the resulting topology of the Fermi surface 

>From the two different ways to define the Fermi surface, i.e.\ via
$|\grad n(\vec k)|$ and inspection of a constant energy scan 
$A(\vec k,\omega=0)$ the latter turned out to be the more precise.
The constant energy scans are able to distinguish the situation where
a peak in the spectral crosses the Fermi surface from that where
it only approaches it.  Thus they were free of spurious bifurcations 
observed in the $|\grad n(\vec k)|$ plots. From the
constant energy scans of the spectrum at the Fermi energy we find that the
Fermi surface changes its topology compared to the non-interacting one as a
function of decreasing doping. While the latter is closed around the zone
center for every finite doping, the interacting Fermi surface at low doping, 
$\delta=0.05$, is hole-like, closed around $M=(\pi,\pi)$. Moreover, the form
and shifts present in the Fermi surface must be taken as clear evidence for
a violation of Luttinger's theorem. Non Fermi liquid behavior is also evidenced
by the spectrum. A strong particle-hole asymmetry is found at the Fermi 
surface crossing near $(\pi/2,\pi/2)$ with hole-like excitations 
having much longer life-times than electronic excitations. 
Furthermore the corresponding self-energy shows strong $\vec k$-dependence 
rendering local approximations like the DMFA irrelevant. Additional structures
in the self energy and a rather large residual scattering rate can also be
interpreted as signs for non Fermi liquid behavior. However, to really
distinguish a non Fermi liquid from a Fermi liquid with a possibly extremely
small energy scale
much lower temperatures must be studied.

With increasing doping the Fermi surface bifurcates at $(\pi,0)$ around a
doping $\delta=0.10$ and from the topology around $(\pi/2,\pi/2)$ one can
infer a tendency towards restoration of Luttinger's theorem and conventional
Fermi liquid behavior. Finally when $\delta=0.20$ the Fermi surface is
electron-like again, i.e. closed around the zone center, and coincides with
the non-interacting Fermi surface. The corresponding spectra show well
defined and sharp quasi-particle peaks around the Fermi energy with
particle-hole symmetry being recovered in the low-energy region.

Although the Hubbard model surely presents an oversimplification of the real
cuprates, the common belief is that at least the essential qualitative
features of the low energy dynamics should be reproduced. In particular, the results presented here show
some interesting qualitative agreement with experiments regarding the behavior
of the different structures observed in the spectra. Moreover, these features
could be related to at least an apparent violation of Luttinger's theorem
or possibly even non Fermi liquid behavior at low doping. From
our results it becomes very clear that neither weak-coupling treatments nor
local theories like the DMFT are able to capture the essentials of the physics
of the cuprate in the weakly doped regime. There are, of course, a variety of
further questions to be addressed. For example, most of the interesting and
controversial ARPES results are for the system Bi2212, which actually is a
bilayer system. Moreover, the simple nearest-neighbor tight binding band
structure used in this paper is definitely not sufficient to describe the
cuprates. Thus, a detailed study about the influence of a finite and negative
$t'$ and also a bilayer coupling on the topology of the Fermi surface or more
generally on the low-energy single particle dynamics is definitely necessary.
We believe that such an investigation can also address some of the still
mysterious features in behavior of the Fermi surface topology of the high-$T_c$
cuprates. Moreover it is very important to find new methods to solve the DCA
self consistency cycle at lower temperatures or preferably at $T=0$. This would
enable us to clearly distinguish between a strong coupling Fermi liquid and
genuine non Fermi liquid behavior at low doping.

\acknowledgements{
We acknowledge useful conversations with
M.\ Hettler,
C.\ Huscroft,
H.R.\ Krishnamurthy,
M.\ Sigrist,
T.M.\ Rice.
This work was supported by NSF grant DMR-0073308 and by the DFG
Graduiertenkolleg ``Komplexit\"at in Festk\"orpern''. 
We acknowledge supercomputer support by the Leibniz Rechenzentrum in Munich
under grant h0301.  This research was supported in part by NSF 
cooperative agreement ACI-9619020 through computing resources provided 
by the National Partnership for Advanced Computational Infrastructure 
at the Pittsburgh Supercomputer Center. }

\end{document}